\documentclass[aps,prl,floatfix,
twocolumn,
superscriptaddress]{revtex4-2}

\usepackage{amsmath,amssymb}
\usepackage{graphicx}
\usepackage{hyperref}
\usepackage{epsfig}
\usepackage{bm}
\usepackage[usenames]{color}

\begin{document}

\title{Andreev Scattering Driven Surface Photogalvanic Effect}





\author{V.~M.~Kovalev}
\affiliation{Rzhanov Institute of Semiconductor Physics, Siberian Branch\\ of Russian Academy of Science, Novosibirsk 630090, Russia}
\affiliation{Novosibirsk State Technical University, Novosibirsk 630073, Russia}



\author{I.~G.~Savenko}
\email[ivan.g.savenko@gmail.com]{}
\affiliation{Guangdong Technion -- Israel Institute of Technology, 241 Daxue Road, Shantou, Guangdong, China, 515063}
\affiliation{Technion -- Israel Institute of Technology, Haifa, 3200003, Israel}


\begin{abstract}
The surface photogalvanic effect is a powerful tool for generating macroscopic direct currents in centrosymmetric film structures via asymmetric scattering of carriers at film boundaries.
Here, we uncover a fundamentally distinct regime of photoinduced transport in normal metal-superconductor heterostructures, in which the macroscopic photocurrent is governed by Andreev scattering at the interface rather than by classical scattering. 
We demonstrate that the interplay between asymmetric surface scattering and Andreev retroreflection yields a giant, strongly frequency-dependent anomalous photocurrent. 
Strikingly, the electron-hole conversion at the normal-superconducting interface dictates a macroscopic sign reversal of the photogalvanic current across the superconducting transition, and induces non-analytic spectral kinks exactly at the pairing gap frequency. 
Our results establish the Andreev reflection-driven surface photogalvanic effect as a platform for optical control of superconducting correlations and suggest a direct, non-invasive spectroscopic probe for gap dynamics in hybrid quantum devices.
\end{abstract}

\maketitle


The investigation of photoinduced transport in hybrid normal metal–superconductor (NS) heterostructures constitutes a rapidly evolving frontier in mesoscopic physics and quantum optoelectronics. 
This field is driven by the imperative to achieve ultrafast optical control over the superconducting (SC) state, a crucial capability for the realization of terahertz detectors, photonic quantum valves, and scalable quantum interfaces~\cite{Klapwijk2014, Beck2020}. 
At the microscopic core of these phenomena lies Andreev reflection~\cite{Andreev1964}: the quantum conversion of a normal-metal electron into a Cooper pair, accompanied by the retroreflection of a hole. 
This fundamental coherent process not only governs junction conductance within the standard Blonder-Tinkham-Klapwijk (BTK) formalism~\cite{BTK1982}, but also mediates the generation of SC correlations in the normal metal, known as the proximity effect~\cite{Buzdin2005_RevModPhys, Bergeret2005}.

While traditional analysis of NS junctions relies on quasistationary frameworks, intense optical illumination drives the system into a strongly nonequilibrium regime. 
The foundational physics of such states traces back to Eliashberg~\cite{Eliashberg1970} and Aronov and Spivak~\cite{Aronov1978}, who demonstrated that high-frequency electromagnetic fields can dynamically stimulate superconductivity via quasiparticle redistribution. 
In contemporary mesoscopic systems, the photoresponse is shaped by a complex interplay between nonequilibrium carriers, microwave fields, and Andreev bound states~\cite{Virtanen2020, Park2024_Nature}. 
Within this context, the surface photogalvanic effect (SPGE) -- the generation of a direct electric current in a near-surface layer upon light absorption in the absence of an external DC bias -- emerges as a highly efficient mechanism for photo-to-current conversion. 

\begin{figure*}[!t] \includegraphics[width=1.8\columnwidth]{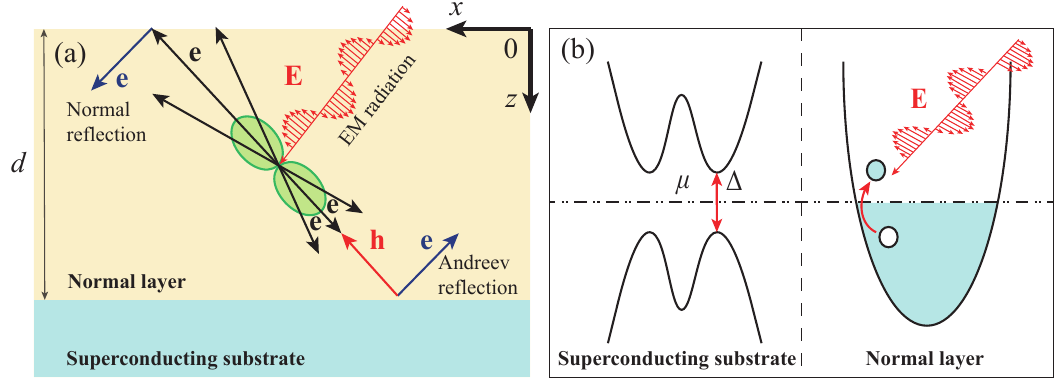} 
\caption{Schematic of the normal-metal--superconductor (NS) heterostructure.
(a) A normal-metal film deposited on a SC substrate is illuminated by obliquely incident electromagnetic radiation. 
(b) Corresponding energy-band diagram illustrating the microscopic subgap transport mechanisms. 
Photoexcited nonequilibrium electrons experience specular scattering at the upper film-vacuum boundary, while coherent charge transfer at the NS interface is dictated by the continuous competition between normal specular reflection (electron-to-electron) and Andreev retroreflection (electron-to-hole).}
\label{Fig1}
\end{figure*}

The microscopic nature of SPGE in metals and semiconductors was discovered by Entin and Magarill~\cite{EM1, EM2}, who proposed a mechanism in which spatial symmetry breaking at the film boundaries induces a macroscopic photocurrent, even when the bulk material has an inversion center. 
This contrasts sharply with the bulk photogalvanic effect, which strictly requires the absence of bulk inversion symmetry~\cite{BelinicherSturman, Sturman1992_Book}. 
Specifically, the asymmetry in electron scattering at the upper and lower film interfaces, coupled with the anisotropic momentum distribution of photoexcited carriers, drives a net in-plane current. 
This SPGE current is highly localized over the mean free path and exhibits a distinct polarization dependence, which has been robustly verified in experiments by Alperovich \textit{et al.} \cite{Alperovich1, Alperovich2}.

Recently, photogalvanic and nonlinear optical phenomena in superconductors have experienced a massive resurgence, motivated by the pursuit of non-reciprocal SC devices~\cite{Nadeem2023, Ando2020_Nature, Parafilo_2025, PhysRevB.110.205413}. 
The microscopic and phenomenological foundations of these effects have been extensively explored~\cite{Melnikov1, PhysRevB.108.L180509, Melnikov2, 9prb-whh8, Melnikov3, Mironov}. 
At the microscopic level, electromagnetic fields induce second-order nonlinear responses, driving phenomena such as photon drag and the ac Hall effect~\cite{Melnikov1}. 
This nonlinearity stems from a local suppression of the order parameter, governed by the interplay between radiation-induced charge imbalance and electron-hole asymmetry. 
In macroscopic transport, such high-order nonlinearities can be captured by generalizing the London equations to include quadratic corrections to the supercurrent-velocity relation. 
This phenomenological framework successfully captures SC diode effects, demonstrating that incident light can establish a net phase gradient across the sample~\cite{Melnikov2}. 
Furthermore, complex optical excitations, such as twisted Bessel beams, enrich this physics by transferring orbital angular momentum to the condensate, dictating unique spatial profiles for photoinduced currents and second-harmonic generation~\cite{Melnikov3, Sonowal_2023}.

Despite this rapid progress in SC optoelectronics, a unified theoretical framework that bridges the macroscopic spatial asymmetry of classical surface scattering and the quantum-coherent nature of Andreev charge transfer has remained elusive. 
A fundamental question arises: How does the quantum mechanical conversion of a photoexcited electron into a retroreflected hole at an NS interface alter the macroscopic directionality and magnitude of the photogalvanic flow?

In this Letter, we resolve this problem by developing a microscopic kinetic theory of the SPGE in NS heterostructures. 
We reveal that the incorporation of Andreev reflection does not merely yield a quantitative correction to the normal-state case, but engenders a radically new optoelectronic response, which, thus, can be called the Andreev scattering-driven SPGE. 
Our analytical framework seamlessly integrates Fuchs boundary conditions for film-vacuum surface scattering with a generalized BTK formalism for the NS interface, yielding the following predictions:
(i) Pure Andreev photocurrent regime: In the specific case of an ideal NS interface and fully diffuse scattering of photoelectrons at the film-vacuum boundary, the photocurrent within the film is entirely determined by the processes of Andreev electron-hole conversion. 
(ii) Anomalous sign reversal: The sign of the Andreev photocurrent is opposite to that of the system's photocurrent in the normal state, i.e., above the critical temperature of the SC transition $T_c$ of the substrate. 
(iii) In the general case of an arbitrary NS boundary potential barrier, the photocurrent sign reversal occurs at a finite light frequency, with the inversion point being dictated by the strength of the potential barrier at the NS interface.
(iv) Gap spectroscopy: The spectral response of the photocurrent exhibits strong non-analytic features (kinks) exactly at the pairing gap edge, offering a purely optical, non-invasive spectroscopic probe for proximity-induced gap dynamics.

By deriving exact closed-form expressions for the photoinduced current in terms of microscopic interface parameters, our results provide a clear roadmap for realizing ultrafast, light-controlled SC valves and advanced terahertz detectors.
%
%
%
%
%
%


\textcolor{blue}{\textit{Kinetic model.}}
Consider a normal-metal film of thickness $d$, occupying the spatial region $0 \le z \le d$, deposited on a SC substrate (Fig.~\ref{Fig1}). 
Let the electrons in the normal metal possess an isotropic parabolic spectrum $\xi_{\bf k}=\epsilon_{\bf k}-\mu = \hbar^2\mathbf{k}^2/2m - \mu$ and obey degenerate Fermi statistics. 
The system is illuminated by an alternating electric field obliquely incident in the $xz$-plane: $\mathbf{E}(x, z, t) = \mathbf{E}_0 e^{-\alpha z/2} e^{i(q_x x - \omega t)} + \text{c.c.}$,
where $\mathbf{E}_0 = (E_{0x}, 0, E_{0z})$ is the complex amplitude, $\omega$ is the radiation frequency, and $\alpha$ is the optical absorption coefficient of the film. 
This external radiation induces a steady-state nonequilibrium correction to the electron distribution function, denoted as $g_{\mathbf{k}}(z)\equiv g(v_x, v_z; \xi_{\bf k}, z)$ with $v_{x(z)}$ the components of the electron velocity. 
The macroscopic direct SPGE current is governed by this correction, which satisfies the linearized kinetic equation~\cite{Zaitsev:2014aa}:
\begin{equation}\label{KineticEq}
   v_z \frac{\partial g_{\mathbf{k}}(z)}{\partial z} + \frac{g_{\mathbf{k}}(z)}{\tau} = G(v_x,v_z;\xi_{\bf k}) e^{-\alpha z},
\end{equation}
where $\tau$ is the momentum relaxation time and $G(v_x,v_z;\xi_{\bf k})$ is the radiation-induced generation rate of nonequilibrium electrons. 

The general solutions of Eq.~\eqref{KineticEq} for electrons moving towards the NS interface ($v_z=|v_z|>0$) and towards the upper vacuum interface ($v_z=-|v_z|<0$) are given, respectively, by
\begin{eqnarray}\label{KineticRight}
  &&g^>(v_x,|v_z|;\xi_{\bf k},z) = C^>e^{-z/l} \\
  \nonumber
  &&~~~~~~~~+\frac{\tau G(v_x,|v_z|;\xi_{\bf k})}{1-\alpha l} \left(e^{-\alpha z}-e^{-z/l}\right),
\end{eqnarray}
and
\begin{eqnarray}
\label{KineticLeft}
  &&g^<(v_x,-|v_z|;\xi_{\bf k},z) = C^<e^{(z-d)/l} \\
  \nonumber
  &&~~~~~~~~~~~+\frac{\tau G(v_x,-|v_z|;\xi_{\bf k})}{1+\alpha l} \left(e^{-\alpha z}-e^{-\alpha d}e^{(z-d)/l}\right),
\end{eqnarray}
where $l=|v_z|\tau$ is the effective mean free path along the $z$-axis, and the integration constants $C^>$ and $C^<$ are determined by the boundary conditions at $z=0$ and $z=d$.

At the upper free surface ($z=0$), we employ the specular-diffuse (Fuchs) boundary condition~\cite{RefBookFuchs}:
\begin{equation}\label{BC_upper}
  g^>(v_x,|v_z|;\xi_{\bf k},0) = D g^<(v_x,-|v_z|;\xi_{\bf k},0),
\end{equation}
where the phenomenological parameter $0 \le D \le 1$ characterizes the specularity of the boundary, with $D=0$ and $D=1$ corresponding to purely diffuse and perfectly specular reflection, respectively. 

At the bottom NS interface ($z=d$), the boundary conditions are governed by the BTK framework~\cite{BTK1982}. 
An incident electron can undergo normal reflection as an electron or Andreev retroreflection as a hole. 
This interfacial scattering can be formalized as:
\begin{eqnarray}\label{NSbondary_full}
&&g^<(v_x,-|v_z|;\xi_{\bf k},d) = B g^>(v_x,|v_z|;\xi_{\bf k},d)\\
\nonumber
&&~~~~~~~~~~~~~~~~~~~~~~~~~~~+A g^>(-v_x,-|v_z|;-\xi_{\bf -k},d).
\end{eqnarray}
Here, $A$ and $B$ are the respective probabilities of Andreev and normal reflection for an incident electron. 
Owing to the symmetry properties of the nonequilibrium distribution function, or more specifically, the fact that the photoinduced correction is odd in momentum, $g^>(-v_x,-|v_z|;-\xi_{\bf -k},d) = -g^>(v_x,|v_z|;\xi_{\bf k},d)$, the boundary condition at the NS interface simplifies to:
\begin{equation}\label{NSbondary_simplified}
  g^<(v_x,-|v_z|;\xi_{\bf k},d) = (B-A) g^>(v_x,|v_z|;\xi_{\bf k},d).
\end{equation}
Matching the general solutions with the boundary conditions yields the constants $C^<$ and $C^>$ (see End Matter).

To proceed, we specify the optical generation source $G(v_x,|v_z|;\xi_{\bf k})$, which drives the anisotropic momentum distribution of photoexcited electrons. 
Evaluating the electron-photon collision integral yields (see the details in End Matter):
\begin{equation}\label{GenerationRate}
\begin{aligned}
G(v_x,|v_z|;\xi_{\bf k}) &= \frac{e^2v_x|v_z|(E_xE_z^*+E_x^*E_z)}{\omega^4}\mathcal{G}_T(\epsilon_{\bf k},\omega), \\
\mathcal{G}_T(\epsilon_{\bf k},\omega) &= \frac{f(\epsilon_{\bf k}-\omega)-f(\epsilon_{\bf k})}{\tau(\epsilon_{\bf k}-\omega)} + \frac{f(\epsilon_{\bf k}+\omega)-f(\epsilon_{\bf k})}{\tau(\epsilon_{\bf k}+\omega)},
\end{aligned}
\end{equation}
where $f(\epsilon)$ is the Fermi-Dirac distribution. Here, $\tau^{-1}(\epsilon_{\bf k})=2\pi N_i|u_0|^2\nu(\epsilon_{\bf k})$ is the energy-dependent momentum relaxation rate, determined by the 3D electronic density of states $\nu(\epsilon)=\sum_{\bf k'}\delta(\epsilon-\epsilon_{\bf k'})$; $N_i$ denotes the impurity density, and the scattering centers are modeled as point-like defects with potential strength $u_0$.

The macroscopic SPGE current density flowing along the film ($x$-direction) is obtained by integrating the nonequilibrium distribution over both the momentum space and the film thickness:
\begin{eqnarray}\label{SPGE_Current}
&&j_x = e \int \frac{d^3\mathbf{k}}{(2\pi)^3} v_x 
\int\limits_0^d dz \left[ g_\mathbf{k}^>(z) + g_\mathbf{k}^<(z) \right].
\end{eqnarray}
%
To evaluate this transport integral, one must determine the reflection probabilities $A$ and $B$ for an arbitrary angle of incidence $\theta$, thus generalizing the BTK formalism. 

In the standard BTK model, interfacial scattering is parameterized by a delta-functional repulsive barrier $H\delta(z-d)$ localized at the NS boundary. 
Consequently, the scattering coefficients depend on both the quasiparticle energy $\xi_{\mathbf{k}}$ and the barrier strength, entering the equations via the dimensionless parameter $Z = H/\hbar v_F$. 
While the original BTK theory assumes normal incidence, the oblique incidence of photoexcited electrons modifies the boundary's effective transparency. 
We account for this by introducing an angle-dependent effective barrier strength: $Z_\theta = Z/\cos\theta$.


\textcolor{blue}{\textit{Thin-film limit.}}
Eq.~\eqref{SPGE_Current} provides a general expression for the photoinduced current density. 
Without loss of generality, we can focus on the thin-film limit, defined by the conditions $\alpha d \ll 1$ and $d/l \ll 1$~\cite{EM1, EM2}. 
In this regime, the total SPGE current simplifies to $j_x = \chi \left(E_x E_z^* + E_x^* E_z\right)$ with
\begin{equation}
\label{SPGEthing}
\chi = \frac{e^3 d^2}{\omega^4} \int \frac{d^3\mathbf{k}}{(2\pi)^3} \, v_x^2 \, \frac{|v_z|\tau}{l} \, \mathcal{G}_T(\epsilon_{\mathbf{k}}, \omega) \left[ \frac{B - A - D}{1 - D(B - A)} \right]
\end{equation}
being the generalized photogalvanic response coefficient. 

Above the SC transition temperature ($T > T_c$), the substrate reverts to the normal state, setting $A = 0$ and $B = Z^2_\theta / (1 + Z^2_\theta) \equiv D_d$, where $0 < D_d < 1$ characterizes the normal-state reflectivity of the $z = d$ interface.
In this limit, Eq.~\eqref{SPGEthing} reduces to $j_x \propto (D_d - D)/(1 - D D_d)$, in full agreement with the classical SPGE current derived for normal-state films~\cite{EM1, EM2}.

A further analytical simplification of Eq.~\eqref{SPGEthing} can be achieved in the low-temperature regime, $0 < T < T_c \ll \mu$. Here, the distribution functions in the generation factor $\mathcal{G}_T(\epsilon_{\mathbf{k}}, \omega)$ reduce to step functions, yielding $f(\epsilon_{\mathbf{k}}-\omega) + f(\epsilon_{\mathbf{k}}+\omega) - 2f(\epsilon_{\mathbf{k}}) \rightarrow \text{sgn}(\xi_{\mathbf{k}}) \theta(\omega^2 - \xi_{\mathbf{k}}^2)$. Furthermore, the momentum relaxation time entering Eq.~\eqref{GenerationRate} can be evaluated at the Fermi energy, $\tau(\epsilon_{\mathbf{k}}\pm\omega) \approx \tau(\mu) \equiv \tau_i$, yielding
\begin{equation}\label{SPGEthing2}
\begin{aligned}
\chi = \frac{e^3 d^2}{\tau_i\omega^4} \int &\frac{d^3\mathbf{k}}{(2\pi)^3} \, v_x^2 \, \text{sgn}(\xi_{\mathbf{k}}) \theta(\omega^2 - \xi_{\mathbf{k}}^2) \\
&\times \left[ \frac{B(\xi_{\mathbf{k}}) - A(\xi_{\mathbf{k}}) - D}{1 - D[B(\xi_{\mathbf{k}}) - A(\xi_{\mathbf{k}})]} \right].
\end{aligned}
\end{equation}
%
In what follows, we use formula~\eqref{SPGEthing2} across distinct transport regimes to elucidate the underlying physical mechanisms.


\textcolor{blue}{\textit{Ideal NS interface: Andreev-driven response.}}
To elucidate the microscopic quantum mechanism of the SPGE mediated by Andreev processes, we first examine an ideal, perfectly transparent NS interface ($Z_\theta = 0$). 
In this clean limit, normal backscattering is completely suppressed in both the subgap ($|\xi_{\mathbf{k}}| < \Delta$) and above-gap ($|\xi_{\mathbf{k}}| > \Delta$) energy regimes ($B = 0$), eliminating the angular dependence of the BTK reflection coefficients~\cite{BTK1982}. 
The macroscopic photocurrent density is evaluated by integrating over the electronic momenta $\mathbf{k}$ in accordance with Eq.~\eqref{SPGEthing2}, yielding:
\begin{equation}\label{SPGE_Andreev}
\chi = -\frac{2e^3 d^2 \nu(\mu)}{3m\tau_i\omega^4} \int\limits_{-\omega}^{\omega} d\xi_{\mathbf{k}} \, |\xi_{\mathbf{k}}| \left[ \frac{A(\xi_{\mathbf{k}}) + D}{1 + D A(\xi_{\mathbf{k}})} \right],
\end{equation}
where $\nu(\mu)$ stands for the 3D electronic density of states at the Fermi energy $\mu$.
\begin{figure}[!t] \includegraphics[width=0.99\columnwidth]{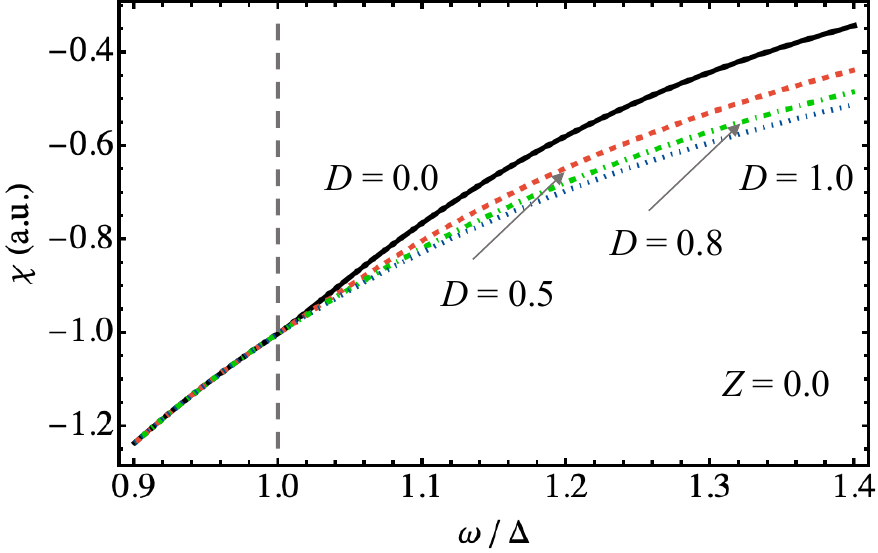} 
\caption{Normalized surface photogalvanic photocurrent coefficient $\chi$ (in arbitrary units) as a function of the reduced frequency $\omega/\Delta$ for an ideal NS interface ($Z = 0$). 
Curves correspond to different values of the specularity parameter $D$ at the film-vacuum boundary ($D = 0$, $0.5$, $0.8$, $1.0$). 
In the subgap regime ($\omega/\Delta < 1$), all curves collapse onto a single line due to unit Andreev reflection probability ($A = 1$). }
\label{Fig2}
\end{figure}

Crucially, for a purely diffusive film-vacuum boundary ($D = 0$), the classical SPGE vanishes identically in normal metals; here, however, a finite SPGE current persists, driven exclusively by Andreev retroreflection. This highlights the intrinsically SC origin of the photoresponse. Another striking feature of the photocurrent in this regime is its sign reversal relative to the normal state (which corresponds to setting $A = 0$, $D = 0$, and $B = D_d$ for $T > T_c$).

\begin{figure*}[!t] \includegraphics[width=1.8\columnwidth]{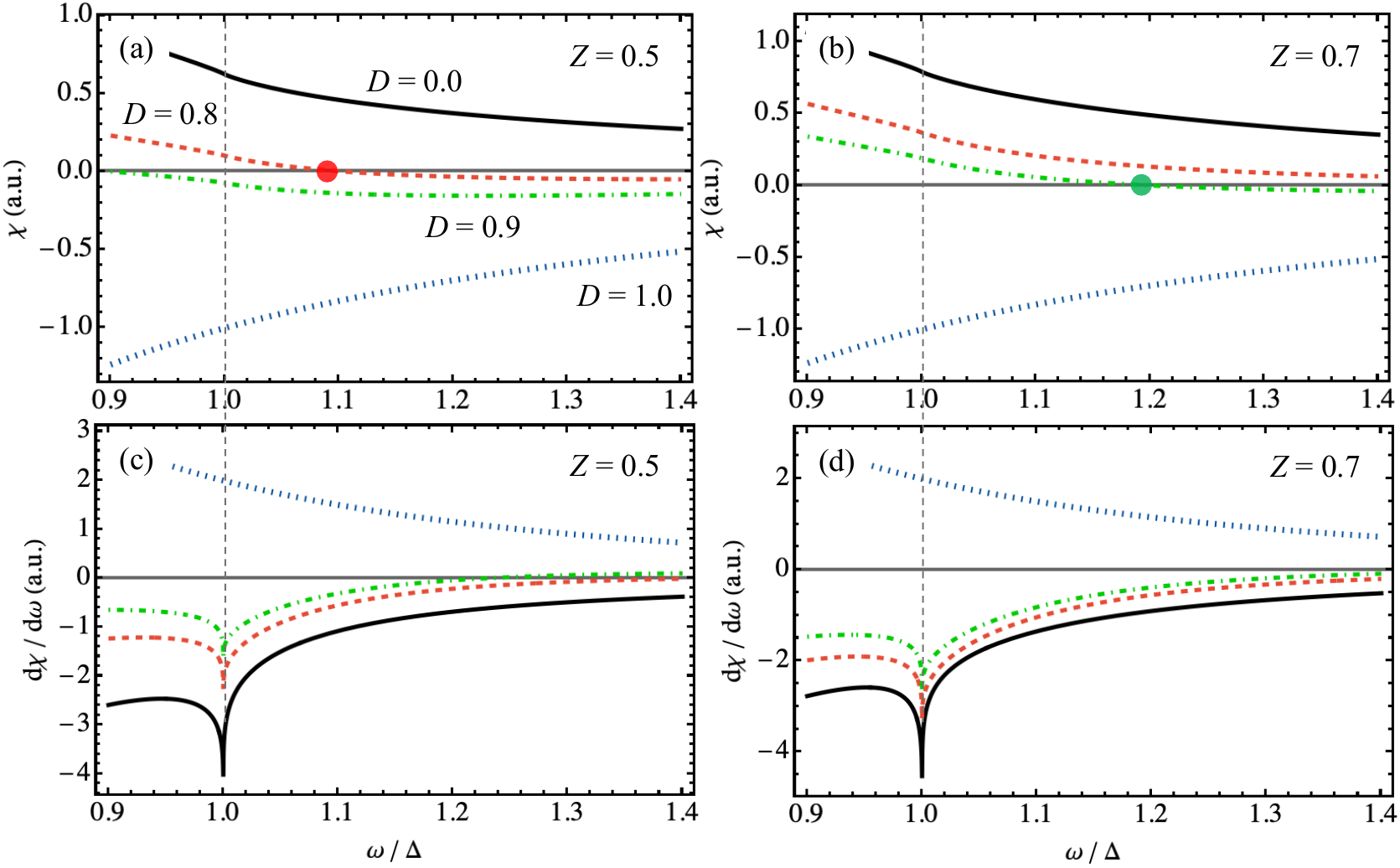} 
\caption{Normalized surface photogalvanic current coefficient $\chi$ and its frequency derivative $d\chi/d\omega$ (in arbitrary units) as functions of the reduced frequency $\omega/\Delta$ for finite barrier strengths at the NS interface. 
The top row shows the photocurrent for (a) $Z = 0.5$ and (b) $Z = 0.7$. The bottom row (c, d) shows the corresponding derivatives. 
Curves denote different values of the film-vacuum specularity parameter $D$ ($D = 0$, $0.8$, $0.9$, $1.0$). 
Red and green dots show the points of electric current sign reversal.}
\label{Fig3}
\end{figure*}

In the purely diffusive limit ($D = 0$), the transport integral in Eq.~\eqref{SPGE_Andreev} can be evaluated analytically, giving:
\begin{equation}\label{SPGE_Andreev2}
\begin{aligned}
\chi &= -\frac{4e^3 d^2 \nu(\mu)}{3m\tau_i\omega^4} \Biggl\{ \frac{\omega^2}{2} + \theta(\omega - \Delta) \\
&\times \Biggl[ \frac{\omega^4}{2\Delta^2} - \frac{\omega^2}{2} - \frac{\omega(2\omega^2 - \Delta^2)\sqrt{\omega^2 - \Delta^2}}{4\Delta^2} \\
&+ \frac{\Delta^2}{4}\ln\left( \frac{\omega + \sqrt{\omega^2 - \Delta^2}}{\Delta} \right) \Biggr] \Biggr\}.
\end{aligned}
\end{equation}
Instead, for a finite $D$ at the film-vacuum boundary ($0 < D < 1$), the integrals in $\chi$ can be evaluated numerically, yielding Fig.~\ref{Fig2}, which demonstrates a decrease of the absolute value of the transport coefficient. 
Most remarkably, in the subgap regime ($\omega < \Delta$), the SPGE current exhibits a complete insensitivity to the parameter $D$.
This behavior arises because the Andreev reflection probability $A(\xi_{\mathbf{k}}) = 1$ below the gap. 
As a consequence, the effective surface scattering factor entering the current expression, $(A(\xi_{\mathbf{k}}) + D) / (1 + D A(\xi_{\mathbf{k}}))$, identically reduces to unity, becoming strictly independent of $D$.

Conversely, in the overgap regime ($\omega > \Delta$), the Andreev reflection probability drops below unity.
Here, the photocurrent becomes highly sensitive to the specularity of the film-vacuum interface. 
Specifically, increasing $D$ enhances the magnitude of the total current due to the constructive contribution of normally specular-reflected electrons at the film-vacuum boundary.


\textcolor{blue}{\textit{Arbitrary barrier strength.}}
The case of an arbitrary barrier strength corresponds to a general transport regime governed by the continuous competition between normal specular reflection and Andreev retroreflection at the interface. 
Formally characterized by a finite BTK parameter $Z \sim 1$, this regime entails a strong energy-dependent suppression of subgap charge transfer. 
Consequently, the photoinduced transport undergoes a continuous crossover from the purely Andreev scattering-driven anomalous response typical of highly transparent interfaces ($Z \ll 1$) to a conventional classical photogalvanic response in the tunneling limit ($Z \gg 1$). 
Below, we capture this crossover by evaluating the transport integrals without invoking any approximations for the parameter $Z$.


Figure~\ref{Fig3} illustrates the calculated SPGE current profiles across different barrier strengths and scattering conditions.
In stark contrast to the ideal barrierless case, the subgap response ($\omega/\Delta < 1$) exhibits a pronounced sensitivity to $D$ due to the energy-dependent suppression of the Andreev reflection probability.
The non-analytic spectral cusp at the gap edge ($\omega = \Delta$) prominently manifests as a sharp discontinuity in the transport derivative, which we demonstrate by plotting the derivative $d\chi/d\omega$ in panels (c) and (d).


\textcolor{blue}{\textit{Thick-film limit.}}
Here, the photoinduced current is governed by the generalized photogalvanic response coefficient:
\begin{eqnarray}\label{thick1} 
\chi &\approx& \frac{e^3 \tau (1 - D)}{\alpha \omega^4} \int \frac{d^3\mathbf{k}}{(2\pi)^3} \, v_x^2 |v_z| \, \mathcal{G}_T(\epsilon_{\mathbf{k}}, \omega).
\end{eqnarray} 
A key feature of Eq.~\eqref{thick1} is its complete independence from the properties of the NS interface. 
Physically, this behavior is straightforward: in a thick film, the incident electromagnetic radiation is fully absorbed within a narrow layer of thickness $\sim 1/\alpha$ adjacent to the film-vacuum boundary. 
Owing to the condition $d \gg l$, momentum-relaxing scattering processes completely randomize the directional motion of photogenerated carriers long before they can reach the NS junction. 
Consequently, the SPGE current in thick films remains entirely decoupled from the underlying SC substrate.

\begin{figure}[!t] \includegraphics[width=0.99\columnwidth]{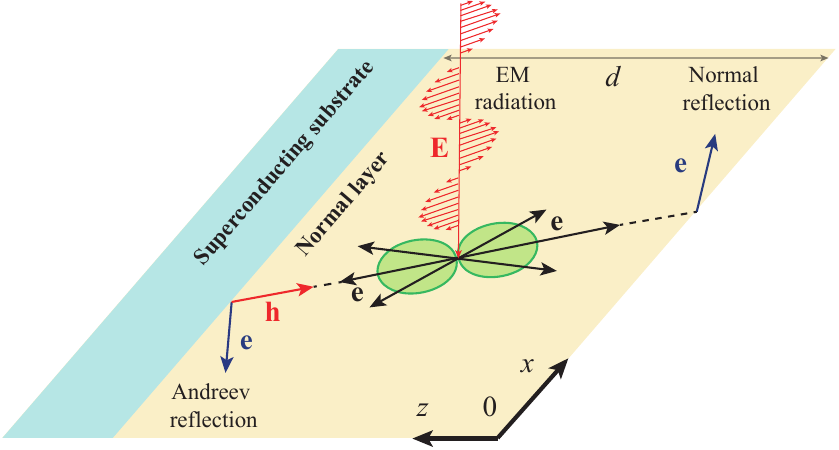} 
\caption{Schematic of the lateral 2D geometry for the surface photogalvanic effect. 
A purely 2D normal-metal strip of width $d$ is laterally coupled to a semi-infinite SC region.
The system is uniformly illuminated by normally incident electromagnetic radiation, with the light's polarization plane oriented at an oblique angle to the NS interface. 
Unlike the quasi-3D thick-film regime, this lateral configuration ensures that the macroscopic photoresponse retains its direct dependence on the quantum Andreev scattering processes regardless of the total strip width.}
\label{Fig4}
\end{figure}

\textcolor{blue}{\textit{Purely 2D systems.}}
The analysis presented above focuses on quasi-3D heterostructures comprising a normal film deposited on a SC substrate.
Remarkably, this Andreev scattering-driven photogalvanic effect can likewise be realized in purely 2D systems. 
There, the photoinduced current is expected to arise in a lateral configuration featuring a 2D normal-metal strip side-coupled to a semi-infinite SC region (Fig.~\ref{Fig4}). 
The effect manifests under normally incident light, provided the polarization plane is oriented at an arbitrary oblique angle relative to the NS interface.

The expression for the SPGE current in this geometry can be derived following the general theoretical framework outlined above. 
In a purely 2D configuration, the normal strip is illuminated uniformly, meaning that the optical absorption profile is spatially constant along the transport direction ($\alpha \to 0$). 
Consequently, by performing the integration over the 2D momentum, the generalized photogalvanic response coefficient takes the form:
\begin{equation}\label{2Dcase} 
\begin{aligned}
\chi &= \frac{e^3}{\omega^4} \int \frac{d^2\mathbf{k}}{(2\pi)^2} \, v_x^2 (l\tau|v_z|) \, \mathcal{G}_T(\epsilon_{\mathbf{k}}, \omega) \\
&\times \frac{(B - A - D)(1 - e^{-d/l})^2}{1 - (B - A)D e^{-2d/l}}.
\end{aligned}
\end{equation} 
In the narrow-strip limit ($d \ll l=|v_z|\tau$), this expression serves as the strict 2D analog of Eq.~\eqref{SPGEthing} derived for the quasi-3D case. 

Notably, even for a wide strip ($d \gg l$), the macroscopic current retains its direct dependence on the Andreev scattering factor $(B - A)$, reflecting the persistent significance of the NS boundary. 
This behavior stands in sharp contrast to the thick-film 3D case: here, the purely 2D geometry and uniform illumination ensure that the incident electromagnetic radiation is continuously absorbed in the immediate vicinity of the NS interface, regardless of the total width $d$ of the normal strip.
%
%
%
%
%
%

While in 3D thick films the photocurrent completely decouples from the superconductor, in the 2D lateral setup the current saturates at wide-strip limits ($d/l \gg 1$) to a $Z$-dependent function that does not depend on $d$. 
Depending on the competition between normal and Andreev reflection encoded in $(B - A)$, the asymptotic 2D photocurrent can even undergo a complete sign reversal as $Z$ varies.
It should be noted that at very small $d/l$, electron-electron (and electron-Andreev hole) scattering begins to play an important role, which changes the transport in the strip.
But the account of these processes is beyond the scope of this manuscript.


\textcolor{blue}{\textit{Experimental observability and bolometric background.}} 
For the proposed Andreev-driven SPGE to serve as a viable spectroscopic probe, two critical experimental conditions must be satisfied: the optical excitation must not destroy the SC state, and the electric current must be distinguishable from trivial thermal effects.
First, the validity of our perturbative kinetic framework requires that the kinetic energy imparted by the electromagnetic field (within an optical cycle) remains well below the pairing gap, $e E_0 v_F / \omega \ll \Delta$. 
For typical superconductors (e.g., Nb or Al with $\Delta \sim 0.2 - 1.5$ meV) probed at sub-THz frequencies ($\omega \sim 0.1 - 0.5$ meV), this restricts the incident field strength to $E_0 \lesssim 10 - 50$ V/cm. 
Such field strengths are readily achievable with standard CW terahertz sources, ensuring the system remains nearly intact by the detrimental effect of the external electromagnetic field (assuring a non-invasive regime). 
Because the magnitude of the derived response coefficient $\chi$ is comparable to that of the normal-state SPGE, we anticipate macroscopic DC currents in the nano- to micro-ampere range, well within the resolution of modern experiments.

Second, intense high-frequency illumination of NS structures inevitably induces quasi-particle heating, generating polarization-independent thermoelectric (Seebeck) currents that can mask delicate optoelectronic phenomena. 
However, the microscopic symmetry of the Andreev-driven SPGE allows for its unambiguous isolation. 
As dictated by our model, the quasi-3D current strictly scales as $(E_x E_z^* + E_x^* E_z)$, while the 2D lateral current depends directly on the oblique angle of the in-plane polarization vector. 
Consequently, the purely thermoelectric background, which scales isotropically with the absorbed intensity $|\mathbf{E}|^2$, can be perfectly subtracted by modulating the polarization state of the incident light.

Furthermore, under oblique incidense, the photon drag effect (PDE) takes place.
Based on the methodology outlined in~\cite{nano11112827}, isolating the SPGE from the PDE requires a rigorous analysis of both the geometric symmetries and the temporal profiles of the generated photocurrents. 
Experimentally, this is achieved by measuring the current in both longitudinal (parallel to the plane of incidence) and transverse (perpendicular) configurations while systematically varying the angle of incidence and the incident light's polarization. 
By mapping the angular, polarization, and temporal dependencies onto the corresponding second-order nonlinear susceptibility tensors, the individual contributions of SPGE and PDE can be disentangled. 


\textcolor{blue}{\textit{In summary,}}
we have established a comprehensive microscopic theory of the Andreev scattering-driven surface photogalvanic effect in normal-metal--superconductor heterostructures. By uniting the kinematics of classical surface scattering with the coherent Blonder-Tinkham-Klapwijk quantum boundary conditions, we demonstrated that macroscopic photoinduced transport is fundamentally dictated by the microscopic competition between conventional specular reflection and Andreev electron-hole conversion.

The most striking consequence of this interplay emerges in the thin-film regime, where the opening of the Andreev channel fundamentally rewires the system's optoelectronic response. For highly transparent junctions, the phase-conjugate nature of Andreev retroreflection triggers a macroscopic sign reversal of the photocurrent compared to the normal state ($T > T_c$). For arbitrary barrier strengths, this electron-hole competition yields a highly tunable photoresponse, in which the direction of the macroscopic current can be inverted simply by varying the incident optical frequency or the interface transparency. 

Crucially, our findings elevate the SPGE from a foundational transport anomaly to a powerful, non-invasive diagnostic tool for mesoscopic physics. Because the lateral photocurrent retains a macroscopic memory of the quantum scattering events at the buried NS boundary, it optically encodes vital information about the superconducting state. The robust, non-analytic cusp in the photoresponse exactly at the pairing gap edge ($\omega = \Delta$) provides a direct mechanism for all-optical spectroscopy of the proximity-induced gap, bypassing the need for invasive transport leads or tunneling contacts that often plague nanoscale device characterization. Furthermore, the extreme sensitivity of the photocurrent's magnitude and sign to the interface barrier ($Z$) offers a unique optical probe for quantifying the specularity and transparency of buried quantum interfaces.

Looking forward, the ability to rapidly modulate the sign and amplitude of the SPGE current via optical excitation frequencies opens compelling avenues for applied quantum optoelectronics. The phenomena predicted here provide a blueprint for realizing ultrafast terahertz detectors, non-reciprocal phase-sensitive switches, and light-controlled superconducting valves operating on sub-picosecond timescales.

We were supported by the Guangdong Basic and Applied Basic Research Foundation under Grant No.~2026A1515012415 and the National Natural Science Foundation of China (NSFC) under Grant No.~W2532001. 
V.M.K. acknowledges the support of the Foundation for the Advancement of Theoretical Physics and Mathematics ``BASIS''. 
We thank Elizaveta Osipova for help with the figures.


\bibliography{biblio}
\bibliographystyle{apsrev4-2}

\begin{widetext}

\appendix

\section{End Matter}

Combining the boundary conditions at the upper and lower interfaces, we find the coefficients:
\begin{equation}\label{ArbitraryConstants1}
\begin{aligned}
C^<(v_x,|v_z|) &= (B-A)\frac{b-aDe^{-d/l}}{1-(B-A)De^{-2d/l}}, 
~~~
C^>(v_x,|v_z|) &= D\frac{b(B-A)e^{-d/l}-a}{1-(B-A)De^{-2d/l}}, 
\end{aligned}
\end{equation}
where we introduced the auxiliary source functions:
\begin{equation}
\begin{aligned}
a &= \frac{\tau G(v_x,|v_z|;\xi_{\bf k})}{1+\alpha l}\left(1-e^{-\alpha d}e^{-d/l}\right), 
~~~
b &= \frac{\tau G(v_x,|v_z|;\xi_{\bf k})}{1-\alpha l}\left(e^{-\alpha d}-e^{-d/l}\right).
\end{aligned}
\end{equation}

Next, we derive the generation rate of photoexcited carriers within the film, induced by an external electromagnetic field. 
A microscopic analysis accounting for photon absorption and emission processes mediated by electron-impurity scattering yields the following expression for the generation rate:
\begin{equation}
\begin{aligned}
G_{\mathbf{k}} &= 2\pi\sum_{\mathbf{k}'}
\Bigl[\langle|M^{(+)}_{\mathbf{k}\mathbf{k}'}|^2\rangle_{\text{imp}}f_{\mathbf{k}'}(1-f_{\mathbf{k}})\delta(\epsilon_{\mathbf{k}'}-\epsilon_{\mathbf{k}}+\omega) 
+ 
\langle|M^{(-)}_{\mathbf{k}\mathbf{k}'}|^2\rangle_{\text{imp}}f_{\mathbf{k}'}(1-f_{\mathbf{k}})\delta(\epsilon_{\mathbf{k}'}-\epsilon_{\mathbf{k}}-\omega)
\Bigr] \\
&\quad - 2\pi\sum_{\mathbf{k}'}
\Bigl[\langle|M^{(-)}_{\mathbf{k}'\mathbf{k}}|^2\rangle_{\text{imp}}f_{\mathbf{k}}(1-f_{\mathbf{k}'})\delta(\epsilon_{\mathbf{k}'}-\epsilon_{\mathbf{k}}+\omega)  + \langle|M^{(+)}_{\mathbf{k}'\mathbf{k}}|^2\rangle_{\text{imp}}f_{\mathbf{k}}(1-f_{\mathbf{k}'})\delta(\epsilon_{\mathbf{k}'}-\epsilon_{\mathbf{k}}-\omega)
\Bigr].
\end{aligned}
\label{appendix1}
\end{equation}
Here, $M^{(\pm)}_{\mathbf{k}\mathbf{p}}$ denote the composite matrix elements describing electronic transitions driven by the combined electron-impurity and electron-photon interaction potentials. The superscripts $(\pm)$ correspond to the absorption and emission of a photon, respectively, while $\langle\dots\rangle_{\text{imp}}$ indicates configuration averaging over the impurity positions. The electron kinetic energy is given by $\epsilon_{\mathbf{k}} = \mathbf{k}^2 / 2m$ (with $m$ being the effective mass), and $\omega$ represents the incident photon frequency. The first two terms on the right-hand side of Eq.~\eqref{appendix1} correspond to the in-scattering (incoming) processes mediated by photon absorption or emission, whereas the remaining two terms describe the out-scattering (outgoing) processes with respect to a given energy state $\epsilon_{\mathbf{k}}$.

In the dipole approximation, configuration averaging of the matrix elements over the random impurity positions yields:
\begin{equation}
\begin{aligned}
\langle|M^{(\pm)}_{\mathbf{k}'\mathbf{k}}|^2\rangle_{\text{imp}} &= \langle|M^{(\pm)}_{\mathbf{k}\mathbf{k}'}|^2\rangle_{\text{imp}} \equiv |M(\mathbf{v}_{\mathbf{k}},\mathbf{v}_{\mathbf{k}'})|^2, \\
|M(\mathbf{v}_{\mathbf{k}},\mathbf{v}_{\mathbf{k}'})|^2 &= N_i u_0^2\left(\frac{e}{\omega^2}\right)^2|\mathbf{E}\cdot(\mathbf{v}_{\mathbf{k}}-\mathbf{v}_{\mathbf{k}'})|^2.
\end{aligned}
\label{appendix2}
\end{equation}
Here, $N_i$ denotes the impurity density, and the scattering centers are modeled as point-like defects characterized by a potential strength $u_0$. Consequently, the generation rate reduces to:
\begin{equation}
\begin{aligned}
G_{\mathbf{k}} &= 2\pi N_i|u_0|^2\left(\frac{e}{\omega^2}\right)^2\sum_{\mathbf{k}'}|\mathbf{E}\cdot(\mathbf{v}_{\mathbf{k}}-\mathbf{v}_{\mathbf{k}'})|^2 \\
&\quad \times [\delta(\epsilon_{\mathbf{k}'}-\epsilon_{\mathbf{k}}+\omega)+\delta(\epsilon_{\mathbf{k}'}-\epsilon_{\mathbf{k}}-\omega)](f_{\mathbf{k}'}-f_{\mathbf{k}}).
\end{aligned}
\label{appendix3}
\end{equation}
Assuming an isotropic electron dispersion $\epsilon_{\mathbf{k}'}$, the interference terms $(\mathbf{E}\cdot\mathbf{v}_{\mathbf{k}})(\mathbf{E}^*\cdot\mathbf{v}_{\mathbf{k}'})$ vanish upon integration. Furthermore, the contribution to $G_{\mathbf{k}}$ arising from the $(\mathbf{E}\cdot\mathbf{v}_{\mathbf{k}'})(\mathbf{E}^*\cdot\mathbf{v}_{\mathbf{k}'})$ terms does not yield a net SPGE current. Consequently, the only non-vanishing terms contributing to the SPGE yield:
\begin{equation}
\begin{aligned}
G_{\mathbf{k}} &= 2\pi N_i|u_0|^2\left(\frac{e}{\omega^2}\right)^2|\mathbf{E}\cdot\mathbf{v}_{\mathbf{k}}|^2  \sum_{\mathbf{k}'} [\delta(\epsilon_{\mathbf{k}'}-\epsilon_{\mathbf{k}}+\omega)+\delta(\epsilon_{\mathbf{k}'}-\epsilon_{\mathbf{k}}-\omega)](f_{\mathbf{k}'}-f_{\mathbf{k}}) \\
&= \left(\frac{e|\mathbf{E}\cdot\mathbf{v}_{\mathbf{k}}|}{\omega^2}\right)^2
\left[\frac{F(\epsilon_{\mathbf{k}}-\omega)}{\tau(\epsilon_{\mathbf{k}}-\omega)}+\frac{F(\epsilon_{\mathbf{k}}+\omega)}{\tau(\epsilon_{\mathbf{k}}+\omega)}\right],
\end{aligned}
\label{appendix4}
\end{equation}
where $F(\epsilon_{\mathbf{k}}\pm\omega) = f(\epsilon_{\mathbf{k}}\pm\omega)-f(\epsilon_{\mathbf{k}})$ and $\tau^{-1}(\epsilon_{\mathbf{k}}) = 2\pi N_i|u_0|^2\nu(\epsilon_{\mathbf{k}})$
with $\nu(\epsilon)=\sum_{\mathbf{k}'}\delta(\epsilon-\epsilon_{\mathbf{k}'})$  the 3D density of electron states. 

Next, we discuss the possible BTK coefficients for different regimes.
The explicit form of the Andreev reflection probability for an ideal (clean) NS contact ($Z=0$) reads:
\begin{equation}\label{BKTClean}
A(\xi_{\mathbf{k}}) = \begin{cases} 
1, & |\xi_{\mathbf{k}}| < \Delta, \\[8pt]
\dfrac{v_k^2}{u_k^2} = \dfrac{|\xi_{\mathbf{k}}| - \sqrt{\xi_{\mathbf{k}}^2 - \Delta^2}}{|\xi_{\mathbf{k}}| + \sqrt{\xi_{\mathbf{k}}^2 - \Delta^2}}, & |\xi_{\mathbf{k}}| > \Delta,
\end{cases}
\end{equation}
where the BCS coherence factors satisfy $u_k^2 = 1 - v_k^2 = \frac{1}{2}\left(1 + \frac{\sqrt{\xi_{\mathbf{k}}^2 - \Delta^2}}{|\xi_{\mathbf{k}}|}\right)$.

Generalizing the standard BTK coefficients~\cite{BTK1982} for an arbitrary effective barrier strength $Z_\theta = Z/\cos\theta$ yields the Andreev reflection probability:
\begin{equation}\label{arbitraryA}
A(\xi_{\mathbf{k}}) = 
\begin{cases} 
\dfrac{\Delta^2}{\xi_{\mathbf{k}}^2 + (\Delta^2 - \xi_{\mathbf{k}}^2)(1 + 2Z_\theta^2)^2}, & |\xi_{\mathbf{k}}| < \Delta, \\[12pt]
\dfrac{\Delta^2}{\Bigl[ |\xi_{\mathbf{k}}| + (1 + 2Z_\theta^2)\sqrt{\xi_{\mathbf{k}}^2 - \Delta^2} \Bigr]^2}, & |\xi_{\mathbf{k}}| > \Delta, 
\end{cases}
\end{equation}
and the corresponding normal reflection probability:
\begin{equation}\label{arbitraryB}
B(\xi_{\mathbf{k}}) = 
\begin{cases} 
1 - A(\xi_{\mathbf{k}}), & |\xi_{\mathbf{k}}| < \Delta, \\[12pt]
\dfrac{4Z_\theta^2(1 + Z_\theta^2)(\xi_{\mathbf{k}}^2 - \Delta^2)}{\Bigl[ |\xi_{\mathbf{k}}| + (1 + 2Z_\theta^2)\sqrt{\xi_{\mathbf{k}}^2 - \Delta^2} \Bigr]^2}, & |\xi_{\mathbf{k}}| > \Delta. 
\end{cases}
\end{equation}
In the limit of a perfectly transparent interface ($Z_\theta = 0$), these relations naturally reduce to the clean-limit expressions~\eqref{BKTClean}.

Furthermore, substituting Eqs.~\eqref{arbitraryA} and~\eqref{arbitraryB} in Eq.~\eqref{SPGEthing2} provides a complete analytical description of the SPGE current valid for any barrier strength. 
Because of the nontrivial angular and energy dependence of these scattering amplitudes, the resulting transport integrals are evaluated numerically.

\end{widetext}

\end{document}